\documentclass[aps,prl,showpacs,twocolumn,superscriptaddress]{revtex4}
\usepackage{bm,color}
\usepackage{graphicx}
\usepackage{amsmath}

\begin{document}
\title {Theory of electromagnon in the multiferroic Mn perovskites \\
--- Vital role of higher harmonic components of the spiral spin order ---}

\author{Masahito Mochizuki}
\email{mochizuki@erato-mf.t.u-tokyo.ac.jp}
\affiliation{Department of Applied Physics, The University of Tokyo,
7-3-1, Hongo, Bunkyo-ku, Tokyo 113-8656, Japan}

\author{Nobuo Furukawa}
\email{furukawa@phys.aoyama.ac.jp}
\affiliation{Department of Physics, Aoyama Gakuin University,
Fuchinobe 5-10-1, Sagamihara, 229-8558 Japan}
\affiliation{Multiferroics Project, ERATO, Japan Science and Technology Agency (JST) 
c/o Department of Applied Physics, The University of Tokyo, Tokyo 113-8656, Japan}

\author{Naoto Nagaosa}
\email{nagaosa@appi.t.u-tokyo.ac.jp}
\affiliation{Department of Applied Physics, The University of Tokyo,
7-3-1, Hongo, Bunkyo-ku, Tokyo 113-8656, Japan}
\affiliation{Cross-Correlated Materials Research Group, RIKEN,
Saitama 351-0198, Japan}

\begin{abstract}
We study theoretically the electromagnon and its optical spectrum (OS) of 
the terahertz-frequency regime in the magnetic-spiral-induced multiferroic 
phases of the rare-earth ($R$) Mn perovskites, $R$MnO$_3$, taking 
into account the elliptical deformation or the higher harmonics of the 
spiral spin configuration, which has been missed so far. 
A realistic spin Hamiltonian, which gives phase diagrams in 
agreement with experiments, resolves a long standing puzzle, i.e., 
the double-peak structure of the OS with a larger low-energy peak
originating from magnon modes hybridized with the 
zone-edge state. We also predict the magnon branches associated with the 
electromagnon, which can be tested by neutron-scattering experiment.
\end{abstract}
\maketitle
The charge dynamics below the Mott gap in Mott insulators
is an issue of intensive recent interest.
The rich structures of the low-energy optical spectrum (OS) are  
associated with the spin degree of freedom in so-called multiferroics, 
which shows both magnetic order and ferroelectricity~\cite{ReviewMF}.
The spontaneous electric polarization in these materials is 
driven by the magnetic ordering, and the strong coupling between the 
dynamics of electric polarization and magnetism is inevitable. 
In particular, in the rare-earth ($R$) perovskite manganites 
$R$MnO$_3$~\cite{Kimura03a}, 
which is the main target of this paper, 
the relativistic spin-orbit interaction
and the spin current play essential roles in 
the multiferroic behavior\cite{Katsura05,Mostovoy06,Sergienko06a}. 
Excitations of the magnet are usually described
as the spin wave or the magnon, i.e., a harmonic 
oscillation of spins around their ground-state configuration. 
It couples to the electric polarization in the multiferroics,
and thus is termed electromagnon~\cite{Smolenski82}.
It is crucial to understand the OS of the electromagnon to design the 
giant magnetoelectric coupling in the terahertz-frequency regime. 

In $R$MnO$_3$, frustration between the spin-exchange interactions
leads to a non-collinear spin spiral in the ground state. 
This simple idea, however, cannot explain the 
rich phase diagrams in the plane of Mn-O-Mn bond angle and temperature, 
and also those under magnetic fields. 
In real systems, there are other interactions originating from the 
relativistic spin-orbit interaction such as the magnetic anisotropy
and the Dzyaloshinskii-Moriya (DM) interaction.
By studying a realistic spin Hamiltonian taking into account 
these interactions, the phase diagrams including the spin-flop 
transition have been understood except for the collinear 
E-type spin phase~\cite{Mochizuki09}. It turned out that
coupling of the spins to phonons given by 
$\mathcal{H}_{sp} = \sum_{ij} g Q_{ij} \vec{S}_i \cdot \vec{S}_j$
plays an important role to stabilize the E-type 
phase, which is expected to play some roles even in 
the neighboring spiral phases.
   
On the other hand, after the first experimental observation of
the terahertz OS in the multiferroic $R$MnO$_3$~\cite{Pimenov06a,Pimenov06b},
it was interpreted as a collective mode corresponding to rotation of 
the spin-spiral plane associated with fluctuations of 
the polarization direction~\cite{Katsura07}. 
Later it turned out that the selection rule and the magnitude of 
oscillator strength ruled out this interpretation~\cite{Kida08,Kida08b}, 
and a new mechanism has been searched for.
The most promising candidate is the conventional magnetostriction
mechanism~\cite{Aguilar09,MiyaharaCD08}, where the electric polarization 
$\bm P$ is given by
\begin{equation}
\bm{P} = \sum_{ij} \bm \Pi_{ij}(\bm S_i \cdot \bm S_j).
\label{eq:PSS}
\end{equation}
Here the vector $\bm \Pi_{ij}$ is nonzero in $R$MnO$_3$
since the inversion symmetry is absent at the center of the Mn-O-Mn bond
because of the orthorhombic lattice distortion and/or the staggered 
$d_{3x^2-r^2}/d_{3y^2-r^2}$ orbital ordering.
This contribution cancels out in the ground state due to the
symmetry, but the dynamical fluctuations of $\bm P$
contribute to the electromagnon excitation. 
Especially, in the non-collinear ground state, the single magnon 
processes at the zone edge originate from Eq.~(\ref{eq:PSS}). 
However, this scenario cannot explain the low-energy peak at $\sim$2-3 meV 
(see insets of Fig.~\ref{Fig02}), 
which is comparable to or even larger than the high-energy peak at 
$\sim$5-8 meV in DyMnO$_3$~\cite{Kida08}, TbMnO$_3$~\cite{Takahashi08}, 
and Eu$_{1-x}$Y$_x$MnO$_3$~\cite{Takahashi09}.
Therefore, the puzzle still remains.

A clue to this issue is the proximity to 
collinear spin phases, i.e. the A-type and E-type spin phases. 
Near the phase boundary, the spin configuration 
is not a simple spiral but is subject to the significant elliptical 
modulation and contains higher 
harmonics~\cite{Arima06}, which is sensitively 
enhanced by the tiny spin-phonon coupling or by the weak magnetic 
anisotropy.
In this paper, we study the role of this higher harmonics on the 
electromagnon excitation, and resolve the long standing
puzzle of the OS in the terahertz-frequency regime.

\begin{figure}[tdp]
\includegraphics[scale=1.0]{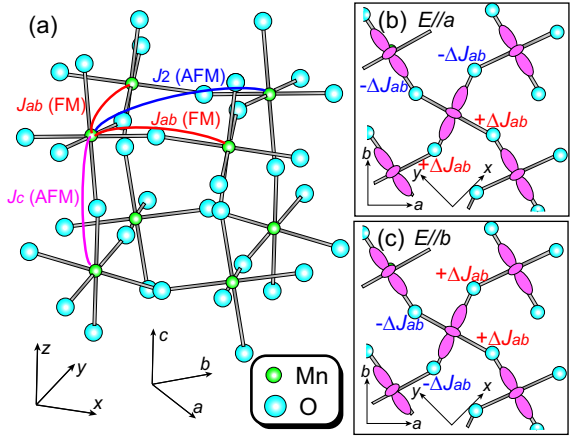}
\caption{(Color online) (a) Superexchange interactions in $R$MnO$_3$
described by the Hamiltonian Eq.~(2). 
(b) Modulations of the in-plane nearest-neighbor ferromagnetic 
exchanges under $\bm E$$\parallel$$a$. (c) Those under $\bm E$$\parallel$$b$.
Here FM and AFM denote ferromagnetic and antiferromagnetic exchanges,
respectively.}
\label{Fig01}
\end{figure}
We start with a realistic spin model for the Mn 
perovskites~\cite{Mochizuki09}. 
The model is basically a classical Heisenberg model on a cubic 
lattice, which contains the frustrating spin exchanges, the DM interaction, 
the single-ion spin anisotropies, and the biquadratic 
interaction.
The Mn spins are treated as classical vectors. 
The Hamiltonian consists of five terms as
$
\mathcal{H}=\mathcal{H}_{\rm ex}
+\mathcal{H}_{\rm sia}^D+\mathcal{H}_{\rm sia}^E
+\mathcal{H}_{\rm DM}+\mathcal{H}_{\rm biq},
$ with
\begin{eqnarray}
\mathcal{H}_{\rm ex} &=&\sum_{<i,j>} J_{ij} \bm S_i \cdot \bm S_j, \\
\mathcal{H}_{\rm sia}^D &=& D \sum_{i} S_{\zeta i}^2,\\
\mathcal{H}_{\rm sia}^E &=&
E\sum_{i}(-1)^{i_x+i_y}(S_{\xi i}^2-S_{\eta i}^2),\\
\mathcal{H}_{\rm DM}&=&
\sum_{<i,j>}\bm d_{ij}\cdot(\bm S_i \times \bm S_j),\\
\mathcal{H}_{\rm biq}&=&-B_{\rm biq}\sum_{<i,j>}^{ab}(\bm S_i \cdot \bm S_j)^2,
\end{eqnarray}
where $i_x$, $i_y$ and $i_z$ represent coordinates of the $i$-th Mn ion with 
respect to the cubic $x$, $y$ and $z$ axes. For the $a$, $b$ and $c$ axes, 
we adopt the $P_{bnm}$ notation [see Fig.~\ref{Fig01}(a)].
The first term $\mathcal{H}_{\rm ex}$ describes the superexchange 
interactions as shown in Fig.~\ref{Fig01}(a). 
The frustration between ferromagnetic $J_{ab}$ and 
antiferromagnetic $J_2$ in the $ab$ plane results in a spiral spin order, 
while the inter-plane antiferromagnetic $J_c$ causes a staggered stacking 
along the $c$ axis. 
The terms $\mathcal{H}_{\rm sia}^D$ and $\mathcal{H}_{\rm sia}^E$ stand for 
the single-ion anisotropies. 
$\mathcal{H}_{\rm sia}^D$ makes the magnetization along the $c$ 
axis hard. On the other hand, $\mathcal{H}_{\rm sia}^E$ causes 
an alternate arrangement of the local hard and easy magnetization axes 
in the $ab$ plane due to the staggered orbitals. 
The term $\mathcal{H}_{\rm DM}$ denotes the DM interaction. 
The DM vectors $\bm d_{ij}$ associated with Mn-O-Mn bonds are expressed 
using five DM parameters as given in Ref.~\cite{Solovyev96}
because of the crystal symmetry; $\alpha_{ab}$, $\beta_{ab}$ and 
$\gamma_{ab}$ for $\bm d_{ij}$ on the in-plane bonds, 
while $\alpha_c$ and $\beta_c$ for $\bm d_{ij}$ on the inter-plane
bonds. The last term $\mathcal{H}_{\rm biq}$ represents 
the biquadratic interaction working between adjacent two spins 
in the $ab$-plane, which originates from the spin-phonon 
coupling~\cite{Kaplan09}. 
We have microscopically determined the parameter values~\cite{Mochizuki09}.

We perform calculations using two sets of the model parameters (A and B) 
as (A) $J_{ab}$=$-$0.74, $J_2$=0.74, $J_c$=1.2, ($\alpha_{ab}$,$\beta_{ab}$,$\gamma_{ab}$)=(0.1, 0.1, 0.16), ($\alpha_c$,$\beta_c$)=(0.4, 0.1), $D$=0.24, $E$=0.3, and $B_{\rm biq}$=0.025, and (B) $J_{ab}$=$-$0.7, $J_2$=0.96, $J_c$=1.0, ($\alpha_{ab}$,$\beta_{ab}$,$\gamma_{ab}$)=(0.1, 0.1, 0.12), ($\alpha_c$,$\beta_c$)=(0.45, 0.1), $D$=0.2, $E$=0.25, and $B_{\rm biq}$=0.025.
Here the energy unit is meV.
These parameter sets give the $ab$-plane spin spiral propagating along the $b$ 
axis with six times periodicity ($q_b$=$\pi$/3) and 
the $bc$-plane one with five times periodicity ($q_b$=2$\pi$/5), respectively.
The former spin state resembles the $ab$-plane spiral in 
Eu$_{1-x}$Y$_x$MnO$_3$ ($x$=0.45) with $q_b$$\sim$0.3$\pi$, while
the latter resembles the $bc$-plane spiral in DyMnO$_3$ with 
$q_b$=0.39$\pi$. 
Note that we adopt the commensurate spin states for convenience of the 
finite-size calculations, whereas the actual spin states are incommensurate. 
However, conclusions of this paper are never affected by this difference. 
Sizes of the systems used are 20$\times$20$\times$6 and 18$\times$18$\times$6 
for respective cases, which match the periodicities of spirals.

We study the electromagnon and magnon excitations by numerically solving the 
Landau-Lifshitz-Gilbert equation using the fourth-order Runge-Kutta method. 
The equation is given by
\begin{equation}
\frac{\partial \bm S_i}{\partial t}=-\bm S_i \times \bm H^{\rm eff}_i
+ \frac{\alpha_{\rm G}}{S} \bm S_i \times \frac{\partial \bm S_i}{\partial t},
\label{eq:LLGEQ}
\end{equation} 
where $\alpha_{\rm G}$(=0.1-0.2) is the dimensionless Gilbert-damping 
coefficient.
We derive an effective local magnetic field $\bm H^{\rm eff}_i$ acting 
on the $i$-th Mn spin $\bm S_i$ from the Hamiltonian $\mathcal{H}$ as
$\bm H^{\rm eff}_i = - \partial \mathcal{H} / \partial \bm S_i$.
Considering the strong reduction of the Mn magnetic moment revealed by 
neutron-scattering experiment~\cite{Arima06}, 
we set $|\bm S|$=1.4.

For the origin of the electromagnon excitation, we consider the coupling 
$-\bm E \cdot \bm P$ between the external electric field $\bm E$ and the 
spin-dependent electric polarizations $\bm P$ given by Eq.~(\ref{eq:PSS}). 
Noticeably this coupling effectively modulates the nearest-neighbor 
ferromagnetic exchanges in the $ab$ plane from 
$J_{ab} \bm S_i \cdot \bm S_j$ 
to $(J_{ab}+\bm E \cdot \bm \Pi_{ij}) \bm S_i \cdot \bm S_j$. 
More concretely, the application of $\bm E$$\parallel$$a$ 
[$\bm E$$\parallel$$b$] corresponds to modulations of the 
in-plane spin exchanges shown in Fig.~\ref{Fig01}(b)
[Fig.~\ref{Fig01}(c)].

\begin{figure}[tdp]
\includegraphics[scale=1.0]{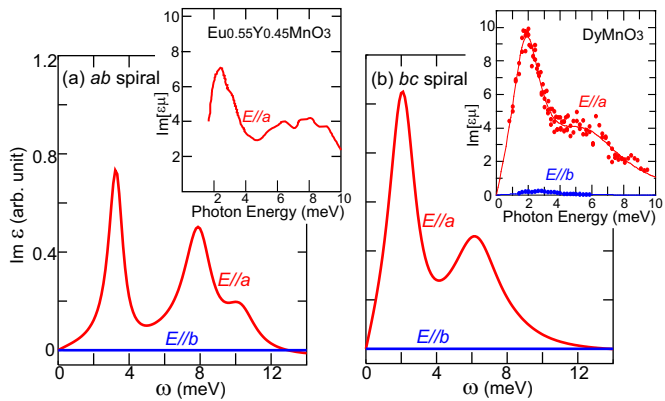}
\caption{(Color online) Calculated electromagnon optical spectra for 
(a) $ab$-plane spiral state ($q_b$=$\pi$/3) with parameter set A,
and (b) $bc$-plane spiral state ($q_b$=2$\pi$/5) with parameter set B.
Values of $\alpha_G$ used for the calculations are 0.1 and 0.2 for 
(a) and (b), respectively.
Insets show the experimental spectra of 
Eu$_{0.55}$Y$_{0.45}$MnO$_3$~\cite{Takahashi09}
and DyMnO$_3$~\cite{Kida08}, respectively.}
\label{Fig02}
\end{figure}
We apply the electric field $\bm E$$\parallel$$a$ or $\bm E$$\parallel$$b$
as a short pulse at $t$=0.
Then we trace the time evolution of $\bm P$ given by Eq.~(1).
The electromagnon spectrum, Im $\varepsilon(\omega)$, is calculated from 
the Fourier transformation of $\bm P(t)$.

In Fig.~\ref{Fig02}, we depict calculated electromagnon spectra for the 
cases of (a) $bc$-plane spiral state with parameter set A and (b) 
$ab$-plane spiral state with parameter set B. 
Irrespective of the spiral-plane orientation, large spectral weight emerges 
at low energy when $\bm E$$\parallel$$a$. In contrast, we observe no
response to $\bm E$$\parallel$$b$ for both cases in agreement 
with the experiment~\cite{Kida08b}.
For comparison, we display the experimental spectrum of 
Eu$_{0.55}$Y$_{0.45}$MnO$_3$~\cite{Takahashi09} and that of 
DyMnO$_3$~\cite{Kida08} in the insets. 
For both cases, we obtain fairly good agreement between theory and 
experiment. 
In the following, we discuss the results for the $ab$-plane spiral case. 
Similar discussion can be repeated for the $bc$-plane spiral case.

\begin{figure}[tdp]
\includegraphics[scale=1.0]{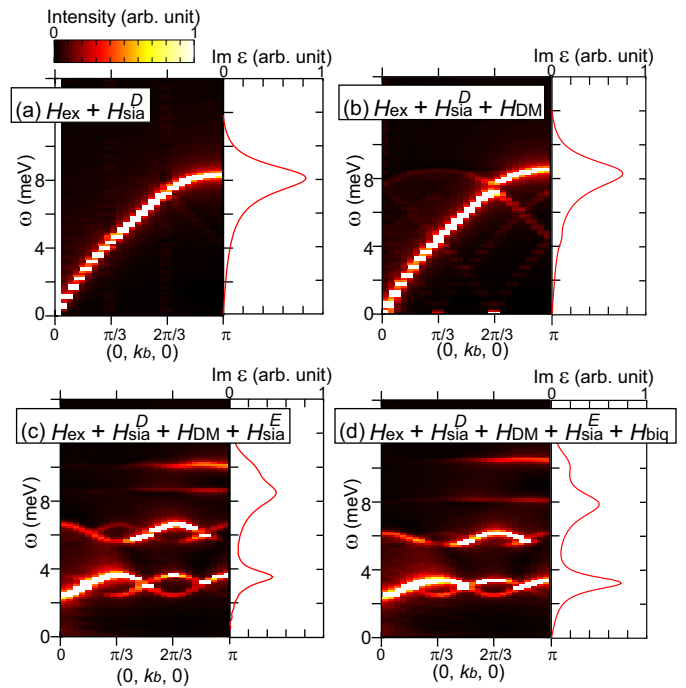}
\caption{(Color online) Calculated intensity map of magnon (left panel) 
and electromagnon OS (right panel) for each Hamiltonian 
with successively adding the interactions. 
(a) $\mathcal{H}_{\rm ex}$+$\mathcal{H}_{\rm sia}^D$ giving a 
spin spiral with a uniform rotation angle. 
(b) Adding $\mathcal{H}_{\rm DM}$ induces negligibly small changes.
(c) Incorporation of $\mathcal{H}_{\rm sia}^E$ causes the elliptical 
deformation or higher harmonics of the spin spiral, resulting in the 
magnon foldings and evolution of the lower-energy peak in the OS.
(d) For full Hamiltonian including also $\mathcal{H}_{\rm biq}$, 
the lower-energy peak in the OS is further enhanced.}
\label{Fig03}
\end{figure}
\begin{figure}[tdp]
\includegraphics[scale=1.0]{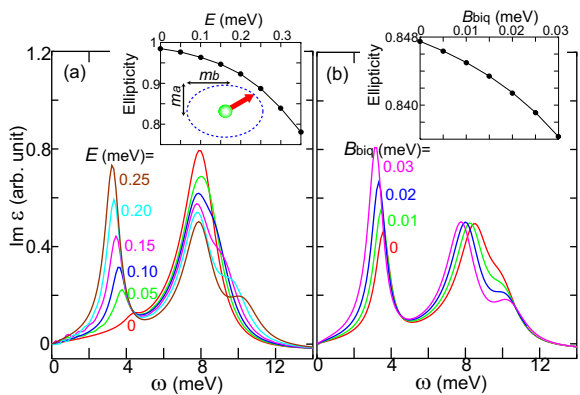}
\caption{(Color online) Calculated electromagnon spectra for various 
strength of (a) single-ion anisotropy $E$ and (b) biquadratic interaction 
$B_{\rm biq}$. Insets show calculated ellipticity of the $ab$-plane spin 
spiral ($m_a$/$m_b$) as functions of $E$ and $B_{\rm biq}$, respectively.}
\label{Fig04}
\end{figure}
To identify origin of the two-peak structure of the electromagnon spectrum,
we calculate electromagnon spectra and magnon-dispersion spectra 
along $\bm k=(0,k_b,0)$, for various cases of interactions 
(see Fig.~\ref{Fig03}).
The magnon spectra are calculated from the Fourier transformation of the 
space- and time-domain simulation data for the spin dynamics 
of $\delta S$$\parallel$($ab$-spiral plane)
after applying $H$ as a short pulse to a single-site spin 
in the $ab$-spiral ground state.
When the Hamiltonian consists of only the spin-exchange term and the 
single-ion anisotropy $D$ term, i.e., $\mathcal{H}_{\rm ex}$ and 
$\mathcal{H}_{\rm sia}^D$ [see Fig.~\ref{Fig03}(a)], 
the spin spiral has a uniform rotation angle. 
In practice, for the case of pure spiral order, translational
symmetry is conserved upon the one lattice-unit translation
if it is accompanied by an appropriate rotation of the spin axes.
Then despite the long-period magnetic structure, 
matrix elements which mix magnon branches do not exist.
In this case, we can see only one peak at a rather high energy of 
$\sim$8 meV corresponding to that in Fig.~\ref{Fig03}(a). 
Incorporation of the DM interaction $\mathcal{H}_{\rm DM}$ 
does not change the spectral shape [see Fig.~\ref{Fig03}(b)]. 

Further adding the single-ion anisotropy $E$ term $\mathcal{H}_{\rm sia}^E$ 
gives rise to folding and anticrossing of the magnon dispersions.
Namely, in the extended Brillouin zone picture,
magnon branches separated by the reciprocal lattice vector 
$\bm G$=(0, $q_b$, 0) are mixed with each other where $q_b$=$\pi$/3 
in the present case.
This gives rise to changes in the spectral shape, i.e. another peak 
of the OS at a lower energy of $\sim$3 meV appears, and the 
higher-lying peak slightly splits into two peaks as shown 
in Fig.~\ref{Fig03}(c). Finally we can see dramatic 
enhancement of the lower-lying peak in the OS
when the Hamiltonian is full including the 
biquadratic term $\mathcal{H}_{\rm biq}$ as shown in Fig.~\ref{Fig03}(d). 
Comparing the electromagnon spectra with the magnon spectra at the 
zone edge, i.e. $k$=(0, $\pi$, 0), we indeed confirm that the 
electromagnon corresponds to the magnon modes hybridized with the 
zone-edge state due to the form factor 
of the magnetoelectric coupling in Eq.~(\ref{eq:PSS})
as described in Ref.~\cite{Aguilar09}. 
Strong magnon anticrossing causes rather flat magnon dispersions.
The magnon branches predicted here would be observed experimentally.

We can see vital roles of the single-ion anisotropy $\mathcal{H}_{\rm sia}^E$
and the biquadratic interaction $\mathcal{H}_{\rm biq}$ 
in Figs.~\ref{Fig04}(a) and ~\ref{Fig04}(b), which show the calculated 
spectra for various values of the coupling constants $E$ and $B_{\rm biq}$,
respectively. Surprisingly the lower-lying peak 
is enhanced strongly by the weak anisotropy or by the tiny 
biquadratic interaction.
In the insets of Figs.~\ref{Fig03}(a) and ~\ref{Fig03}(b), we show 
calculated ellipticity of the spin spiral for each case, which is 
defined as the ratio of amplitudes between the $a$-axis spin component 
and the $b$-axis spin component ($m_a$/$m_b$)~\cite{Yamasaki07a}.
We see that the enhancement of 
the low-lying peak are accompanied by the decrease of ellipticity. 
Note that the elliptical deformation of the spin spiral is more 
significant when the ellipticity is decreased more from unity.
This indicates that the origin of the low-energy peak at 
$\sim$2-3 meV is the magnon-branch mixing due to the 
elliptical modulation of the spin spiral.

Let us now discuss the $R$ dependence of the electromagnon spectra
of $R$MnO$_3$.
In TbMnO$_3$,  the spectrum has more weight at the 
higher-energy peak~\cite{Takahashi08}, while
in DyMnO$_3$ with a smaller ionic $R$-site radius ($r_R$),
the spectral weight is shifted to lower energy~\cite{Kida08}.
In the systematic study of Eu$_{1-x}$Y$_x$MnO$_3$,
transfer of the spectral weight to lower energies
has been reported as the Y content $x$ is increased 
(or as the averaged $r_R$ is decreased)~\cite{Takahashi08}.
Note that with decreasing $r_R$,
we approach the E-type phase, which results in a stronger influence of the
biquadratic interaction.
So far, the magnon spectrum in $R$MnO$_3$ with a spiral spin order 
has only been reported for TbMnO$_3$~\cite{Senff07}.
In the compounds closer to the E-type phase,
we expect that the magnon spectrum exhibits significant foldings.
Further experimental data for these compounds are necessary to
examine our theory.

In some of the $R$MnO$_3$ compounds, deviation of frequencies between 
the magnetic resonance and the electromagnon has been reported~\cite{Kida09}.
Within our theory, the frequency of the 
magnetic resonance at the zone center $\bm k$=0
coincides with that of the electromagnon at $\bm k$=(0, $\pi$, 0) only if
they are connected with a multiple of $\bm G$=(0, $q_b$, 0).
When the spiral magnetic order is incommensurate,
magnetic resonance and electromagnon absorption occur at 
different points in the Brillouin zone. Thus frequencies of 
these two spectra do not necessarily coincide with each other.

In summary, we have studied the electromagnon dynamics
taking into account all the relevant interactions and anisotropies in 
the spin Hamiltonian for $R$MnO$_3$. It is found that the
elliptical modulation or higher harmonics of 
the ground-state spiral spin configuration has crucial 
influences on the foldings of the magnon dispersion and also
the electromagnon spectrum, which resolves the long standing 
puzzle of the low energy peak in the OS.

The authors are grateful to N. Kida, S. Miyahara, Y. Takahashi,
J. S. Lee, and Y. Tokura for fruitful discussions.
This work was supported in part by Grant-in-Aids 
(Grant No.19048015, No.19048008, No.21244053 and No.17105002) and 
NAREGI Nanoscience Project from the Ministry of Education, 
Culture, Sports, Science, and Technology.


\end{document}